# On the economy of web links:

# Simulating the exchange process

Boris Galitsky and Mark Levene

School of Computer Science and Information Systems
Birkbeck College, University of London
Malet Street, London WC1E 7HX, UK
{b.galitsky, m.levene}@dcs.bbk.ac.uk

In the modern web economy hyperlinks have already attained monetary value as incoming links to a web site can increase its visibility on major search engines. Thus links can be viewed as investment instruments that can be the subject of an exchange process. In this study we build a simple model performed by rational agents, whereby links can be bought and sold. Through simulation we achieve consistent economic behaviour of the artificial web community and provide analysis of its micro- and macro-level parameters. In our simulations we take the link economy to its extreme, where a significant number of links are exchanged, concluding that it will lead to a winner take all situation.

## Introduction

The World Wide Web is known to be scale-free: it includes a few nodes, or web sites, that have many links, and many nodes having only a few links (Barabasi 2003); this gives rise to a power law distribution on incoming links (Broder et al 2000). The formation of links is explained through the process of 'preferential attachment', where sites having more incoming links are more likely to be linked to than sites having less incoming links, leading to the 'rich get richer' phenomenon. This explanation does not consider how business practices and other forms of human activity may effect the distribution of incoming links, with the aim of gaining some commercial or other advantage. In this paper we model the process of establishing web links through the behaviour of rational agents on the web who sell and buy links in an attempt to accumulate resources as a result of the link exchange process.

It has been recently pointed out in the hypertext literature that links have become the currency of the web (Walker 2002). The value of a link to a site is measured by its capacity to improve the visibility of the site on web search engines in terms of attaining high ranking (preferably in the top ten hits) for certain keywords related to the site's business. The regulation of the value of links is controlled by the search engine's algorithms, the most pronounced example being Google's PageRank (Brin and Page 1998). Google, and other web search engines, who interpret a link to a web site as a vote for that web site by the author of the link, so the more votes a web site has the more important it is considered. Google also takes into account the importance of the source of a link, so a link from a more prominent site is worth more than a link from a less prominent one. Since web search engines have become the gatekeepers of the web, visibility of a site through web searches has become an essential ingredient for the survival of the site. This is the reason why webmasters may be willing to participate in link exchange programs such as link farms and web rings, although there is a risk involved, since search engines may consider this activity as trying to unfairly manipulate their search engine ranking and will retaliate by dropping their importance to a low level.

In spite of the certain studies addressing the economic issues of a link exchange economy, we have not found any attempt to simulate this process, or to try and predict the consequences of this exchange process becoming the norm rather than an exception. A related approach to modeling economic behaviour is, for example, the massive multi-player online game Everquest (Castronova 2001), which demonstrates that virtual worlds can have real economies.

We provide a simple model that describes the process of establishing a link, which is independent of the content of the site, that is, our model does not take the content of a site into consideration when linking to a site. In our model the reason why such links may be established is that the creator of the link receives a monetary value in return for increasing the value of the web site linked to. As a consequence the web site that purchased the link can hope to sell links from its site to other for a value which is increased as a result of the exchange. On the other hand, as a result of the transaction the web site that sold the link has more buying power to increase its site's value even further.

Our economy-motivated model for the web topology is related to the graph-theoretic approaches to understanding the web topology (Barabasi 2003, Faloutsos 1999). Analysing and modelling the web topology is of a significant practical interest, since knowledge of the network's topological properties enables researchers to optimise network applications and to conduct more representative simulations of how the web is evolving (Levene et al. 2002).

It has been discovered (Zhou and Mondragon 2003) that large, well-connected nodes have more links to each other than to smaller nodes, and smaller nodes also have more links to the larger nodes than to each other. This is in agreement with the model we are suggesting in this paper, since the choice of whom to buy from is preferential (we prefer to buy a more expensive items of better quality) and there is a 'winner take all' situation in the sense that most sites will prefer to link to larger sites. The findings from such work could contribute to better strategies for optimising network traffic flow, network reliability and security, and building network topology simulators.

The rest of the paper is organised as follows. We discuss the phenomenology of the web economy, observe experimental data and discuss the assumptions we make in our simulation of the link exchange process in a step-by-step fashion. Then we present the numerical simulation, starting from a closed set of web agents, who proceed to exchange links, and are joined by new agents who are introduced into the simulation as the game proceeds. We then justify our selection of the most realistic simulation parameters, and perform a comparison with the current web economy. Finally, we hypothesise on how the link exchange process may look like in the future.

# Distribution of links and wealth on the web

The fact that large, well-connected nodes have more links to each other than to smaller nodes, and smaller nodes have more links to the larger nodes than to each other, has been recently numerically analysed (Zho and Mondragon 2003). Their results show that 27 percent of connections are among the largest 5 percent of nodes, 60 percent connect the remaining 95 percent to the largest five percent, and only 13 percent of connections are between nodes not in the top 5 percent. These observations suggest that the dynamics of internet links tends towards the growth of the larger nodes in a more accelerated and concentrated manner than was previously thought.

Our model is based on the currently available data, which exhibits the phenomenon of 'the rich get richer' or in some cases the more extreme 'winner takes all' situation, both for incoming web links and accumulated resources of business web agents. (This phenomenon may depend on the subset of web sites considered within a selected industry sector.) Clearly, the number of links to the web site of an agent is strongly correlated with the resultant accumulated resources, and vice versa, since the higher the agent's resources, the higher the number of links it is capable of attaining or in our case buying.

There is a strong constraint on the mutual dependence of web links and resources that is based on the rationality of agents' behaviour. Why would an agent start a web link-based business if it knows that it will ultimately lead to a 'winner takes all' situation? An agent should be motivated by the market conditions to invest its resources in web links so that it would be able to get a return on investment by selling links from itself to other sites. Such expectations of an agent should be backed up by the dynamics of link exchanges. In particular, an agent should join the link exchange process at the stage when it is profitable for everyone to do so because of the certainty that other agents will join the game in the future. Agents joining the game later are forced to continue playing the link exchange game, since quitting would lead to the loss of their initial investment. The choice is either to lose their current assets or to take advantage of the possibility to win in the future.

Therefore, the link exchange process can be considered from the perspective of investment. However, such an investment instrument as web links possesses the following peculiarities:

- Universality: any two agents can establish a link from one to another.

- Specific topology: the network of links between investors and investment recipients can be parameterised by the graph-theoretic properties such connectivity, paths, cliques and diameter.
- A clear and uniform way to estimate the superposition (portfolio or selected combination) of investments (what we call the gain rank). Naturally, the higher the number of links to a seller web site, the higher the value of a link from it to a given site is. Furthermore, not only the number but the values of (second order) links from other web sites to the seller site contribute to the gain rank, which is the total value of a given link.
- Full distributed knowledge: every agent knows the links between other agents, and it is impossible to conceal them. Also, agent are knowledgeable about the gain ranks of each other.

In the third item above we use the notion of gain rank which is a concretisation and simplification of PageRank, the metric used by Google to measure the importance of a site independently of its content or relevance to a given user query. In contrast to the PageRank, the gain rank of an agent's web site, as we define it herein, takes into account only the number of the first and second order links going into a web site.

In our current model we consider only the 'pure web' component of the business model, where the links are bought to advertise a product and service, but we then take into account only the monetary value of selling the links themselves and not the product value. So, for example, if we sell sponsored links from our site, we may boost traffic to our site by buying sponsored links to our site from another agent. We hypothesise that in the real world, the pure web component of interacting businesses can also be considered as a closed and separate entity from the exchange of goods other than web links. We model the market component, where the value of links from a web site depends on the number of incoming links to it, irrespectively of its content, advertised products and brands. The total value of the company's web site would then be the sum of the values of incoming links and the value of non-web assets.

# Web links and search engines

Prior to the wide use of web search engines users were surfing the web with the aid of humanly edited directories, with Yahoo being the predominant one. The first generation of search engines led by InfoSeek and later AltaVista ranked web sites mainly according to their content using traditional Information Retrieval techniques tailored to the web. Search engines who base their ranking purely on content are open to spammers who can modify their content using various techniques such as keyword stuffing, hidden text and cloaking (Thurow 2003), in an attempt to improve their ranking for chosen keywords. Search engines also realised that they could sell keywords to the highest bidder in return for placing these sites high up on their ranking for the keywords bought, resulting in a lucrative business model. To counter this, the growing understanding of web users that search ranking can be manipulated on a commercial basis, reduced the credibility of search engines engaging in such practices (Introna and Nissenbaum 2000).

Then came Google leading the pack of second generation search engines, which incorporate metrics that rank sites according to link analysis that is independent of content, and clearly separating advertising (sponsored links) from search results. Google have raised the bar on search, and have succeeded in attaining an impressive 75% of web searches in August 2003. Using link analysis as a means of ranking web pages and sites is not without controversy, since links are already considered by some as the 'currency of the web'. As mentioned earlier, being visible on search engines is now playing a big part in the business model of many web sites, which means that metrics such as Google's PageRank (Brin and Page 1998) are monitored closely by many webmasters who will be willing to invest cash in order to raise their score. (Webmasters can view their PageRank value on a logarithmic scale of 0-10 by installing the Google toolbar.) Although the PageRank is just one factor out of many that determine Google's ranking for a query, the perception is that having a high PageRank is of prime importance for visibility. A justification for this can be seen when a web site is in competition with many others with similar content, and the fact that web users rarely scan through more than one page of search results. Google strongly discourages practices that try and unfairly manipulate the PageRank value of a site, which has led to a fierce battle that is likely to continue between search engine optimisers and Google.

We observe that this situation is quite delicate for the search engines' business model. On the one hand, to follow the portal revenue model, a keyword search agent may sell links in the same way it sells banner ad impressions. On the other hand, the users must trust PageRank as a measure of quality rather than a means to gain revenues from advertising. The conflict can be resolved by establishing a ratio between the unbiased and commercial components of computing the PageRank score. This solution is somewhat similar to the commercial scheduling in other media, i.e. TV, radio and newspaper, where the ratio has been well established.

With regards to web links as a means of advertising, the ratio between content-based and commercial links is yet to be established. Clearly, the number of sponsored (commercial) links can grow until it reaches a saturation point, when keyword search portals such as Google or Ask Jeeves will be forced to restrict the number of sponsored links to avoid ruining their reputation. In our simulation we can reproduce the ratio between commercial and content links by controlling the time at which link exchange terminates. In the following, when computing the resources and gain rank of agents, we will concentrate solely on commercially motivated links.

# Rationality of the agents and uniqueness of the model

The reader may expect that a wide spectrum of models can reproduce a power law distribution for resources emerging from the process of link exchange. The main constraint that comes into play that allows any statement about the model's uniqueness is the requirement of rationality. Indeed, a trivial model arises when an agent is just assumed as a determined contributor to the 'winner who takes all' scenario.

In particular, in a deterministic model, where an agent is assumed to be an ideal reasoner with prediction capabilities based on complete knowledge of the state of play, the agent would *not* intentionally participate in the link exchange game with the knowledge that its resources would be lost, given that the initial resources determine the outcome. In such a deterministic environment the link exchange process is ultimately profitable only for the agents having an initial advantage, who expect that other agents will knowingly join the game and lose. Therefore, a deterministic strategy cannot bring rational agents to a power law distribution of resources and gain rank, since there are very few winners and the rest are all losers.

Hence there is a conflict between the macro-level conditions for attaining a power law distribution, and the micro-level, which is led by rationality. Rationality imposes a strong set of constraints on the possibility to reconcile these two levels into a coherent strategy for agents formed by link exchange rules.

We are inherently interested in modelling the rational behaviour of agents who join the game that displays real world phenomena such as a power law distribution of resultant resources. The conflict of requirements for *rationality* on the micro-level and a certain *distribution* on the macro-level is expected to deliver a limited class of game strategies. In this paper we show how varying the rate of agents joining the game can deliver strategies which are irrational as well as rational.

# Experimental observations

In the real business world, the distribution of resources (and, sometimes, investment instruments) occurs in accordance to a power law distribution. There are a number of deviations from this observation, when we consider special forms of investment, particularly, R&D contributions in the technology sector, which are rather close to a normal distribution (for example, the cost of production means and business conduct). For example, a variety of web-based businesses may have similar expenditures for web servers, so roughly the same amount of initial investment for domain-independent web infrastructure is required (other start-up expenses, including marketing, vary significantly from one business domain to another). However, the distribution of returns for these investments will still result in power law distribution, leading to a 'winner takes all' situation. We believe that such a situation is common for competitive environments: every participant makes a certain effort, expecting to win, and some of them who put in slightly higher efforts are the winners.

In the current study we intend to model the most general situation, thus we have collected some empirical evidence of companies' worth (including web-based businesses), and data on the total number of incoming links to the company (also including the web-based businesses). We have collected data and obtained the distributions based on different criteria (belonging to a stock market, technology or industry sector). Furthermore, to increase the evidence of the link distribution, we have considered the number of documents that point via a hyperlink to a given company.

As an estimate of the distribution of resources, we collected data for 256 companies from various industry sectors obtained from Yahoo Finance. The companies chosen were the ones with the best stock performance over three months as of May 2003. The histogram in the left-hand side of Figure 1 contains the number of web documents mentioning a given company from the list of the 256 companies for which we obtained the data (companies for which less than 100 documents were found are not shown in the Figure). The histogram in the middle of Figure 1 depicts the distribution of worth in dollars (as indicated in the figure), and the histogram on the right-hand side of Figure 1 shows the histogram for the number of links to the web sites of businesses, randomly selected from the above 256 companies.

We consider all these parameters as indicators of investment returns (resources) rather than investment means (in our case, gain rank); they roughly obey a power law, but there are characteristic areas on the histogram that are inherent to the particular nature of a process. The number of documents and links are the results of marketing efforts whose volume are a reflection of the resources of the above set of companies. Note that this data is presented as background information, providing motivation for our model. We have not as yet analysed how the parameters above evolve in time and whether their values are dynamically correlated.

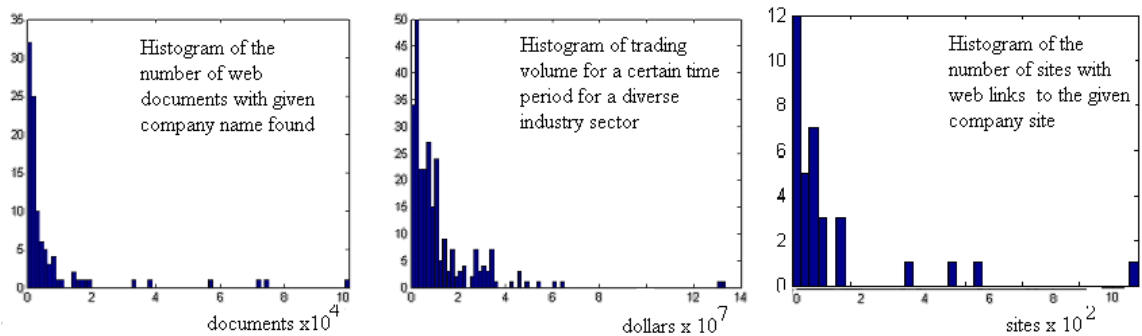

Figure 1 The nature of web-related distributions.

In this study we attempted to follow the train-and-predict methodology in the following way. The initial fragment of our simulation with a low number of commercial links is expected to align our model with the real-world data, and predictions on how the link economy may evolve through the link exchange process will be made by the further simulation steps. The resource distributions attained at the initial steps of our simulation are consistent with power law distributions as observed in Figure 1; however, we are unable to provide experimental measurements for commercialised links because they are still relatively rare and hard to indetify.

# Simulation environment

We have simulated the exchange process between 100 of agents, which have initial resources of 1000 units. To break the symmetry between the agents, initial resources are distributed slightly unevenly, so that an agent joining the game at a later stage has additional resources proportional to the time step when it joined the game. This gives late-comers a fair chance in a game, which in any case has a first mover advantage. The gain rank is computed as the sum of the first-order links and the weighted second order link (with coefficient set to 0.5).

## Pricing and buying strategy

Two components of the link exchange process have to be chosen so that the resulting distribution of resources is a power law for rational agents. These are:

1) For each agent, the mechanism of how to price a link from itself to a buyer's site.
2) For each agent, given its resources, the mechanism of how to optimally choose the price it is willing to pay for a link (and, therefore, from whom to buy the link).

The simplest approach to pricing a link is to make it proportional to the gain rank. In this way the price is equivalent to the objective value of the goods (links) that is accepted by each agent. As to the buying strategy, we choose the simplest, yet rational mechanism, where agents prefer to buy the most expensive link they can afford without spending all their resources. We disallow agents to spend more than half of their resources at each step. Moreover, we constrained the model so that agents can only buy a single link at each step. On the other hand, there is no limitation on the number of links an agent can sell. We may limit this freedom in future simulations to reflect the fact that agents may not want the reputations of their site to drop too rapidly as a result of too many links being sold at any time.

# A simple process

We start with the description of a simulation with a simple model of the exchange process when all agents start simultaneously In Figure 2 we show the results for 30 agents participating in the game. The step that is depicted is at the stage when approximately half of the links have been sold. Resources are distributed around the initial value of 1000 units, deviating not more than 40%. At the same time, the gain rank of all agents is the same; the deviation in resources is caused by the exact number of links each agent has sold. Also, there is no visible order within the link matrix (the right chart on the bottom); the link matrix of all links currently established is a binary matrix, where a cell <i,j> is 1 if there is a link from agent i to agent j and 0 otherwise.

In Figure 2 we show the main indicators of the link exchange process. The charts in the figure present the following information:

1) Resources of agents and the induced histogram (two charts in the top of the figure);
2) Gain rank of agents and the induced histogram (two charts in the middle of the figure);
3) Indication of which agent bought a link from whom at the current step (leftmost chart at the bottom of the figure);
4) The total number of links for each agent (middle chart at the bottom of the figure);
5) The link matrix (right chart at the bottom of the figure).

In the following sections we will use the same set of charts to observe the properties of the exchange process in the case when new agents join the process at different rates.

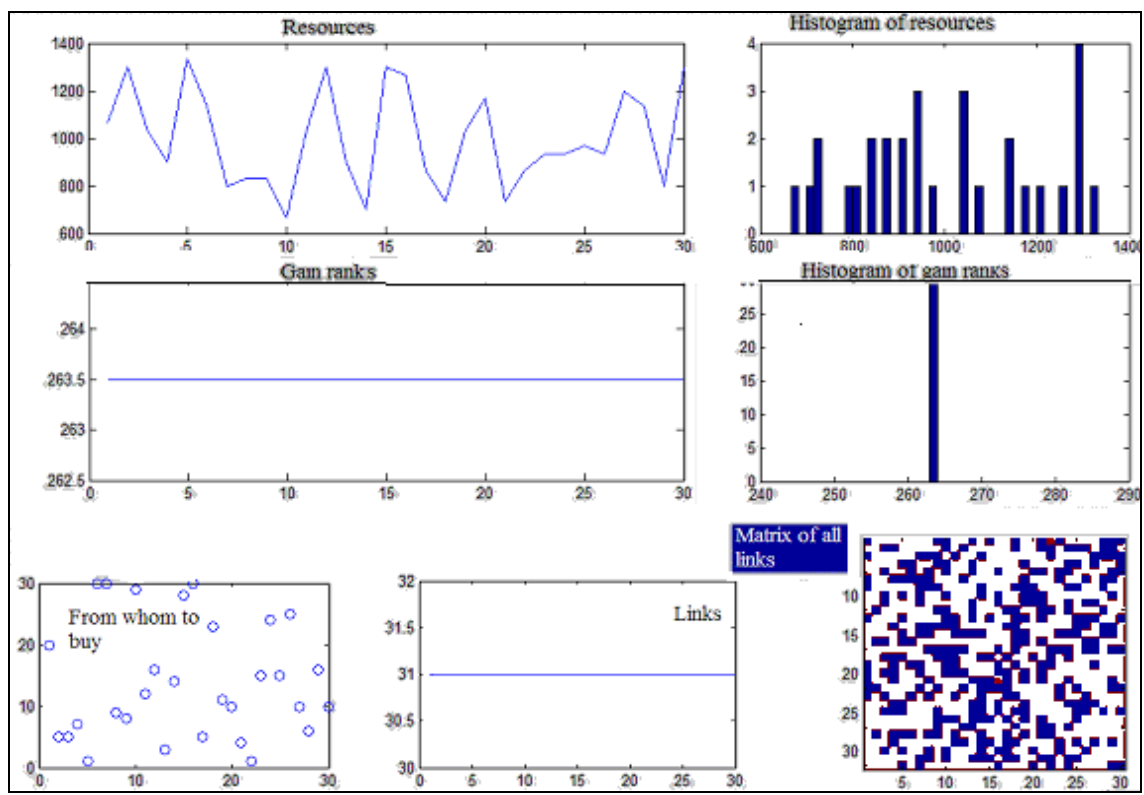

Figure 2: A simple link exchange process where all agents join the game at its initial stage. Note that all agents have the same gain rank.

# A process with joining agents

To develop the reader's intuition concerning the simulation, we present screen-shots showing the simulation step by step. The process, just a few steps after it has started, is depicted in Figure 3a. Agents 1 to10 have joined at the very beginning, and agents 11 to

21 have consequently joined the link exchange process at a later stage. There is a strong deviation of resources for the initial set of agents from 500 to 2400 units, starting from an initial resource of 1000 for each agent. The deviation is due to the discrepancy in the number of links that were sold by each of them.

It is clear from the matrix of all links that initial agents prefer buy from themselves, and that newly joining agent tend first to buy links from the initial set of agents and later from newer agents, whose links are cheaper because they have a lower gain rank. A high number of zero cells on the right side of the matrix shows that links of new agents are less popular than the more expensive links of initial agents; these popular links have higher gain rank. Hence newly joining agents lose their resources at least initially and may gain them later only by selling their links to agents which join the game later than themselves. More expensive links of initial or early agents may be unaffordable for late coming agents, therefore they are forced to buy less expensive links from those agents who joined the in middle game.

Initial agents have similar gain rank, and for the new agents their gain rank grows proportionately to the time since they joined the exchange process. In contrast to the number of links that grows strictly proportionately to the time spent in the game, there are deviations in gain rank values because of the randomness feature of the seller selection process (see chart at the bottom left of Figure 3a). Besides the matrix, we can see than initial agents prefer to buy from themselves, and the later agents buy from the agents which joined earlier.

Regarding the histogram of the resources we see the area for the later agents on the left bar, for 80 agents who have not yet joined by the middle game, and a few small bars for the initial agents on the right. Regarding the gain rank histogram one can see the 80 agent bar on the left and the areas for later agents in the middle and initial agents on the right. This is clearly a 'winner takes almost all' situation for the initial set of agents.

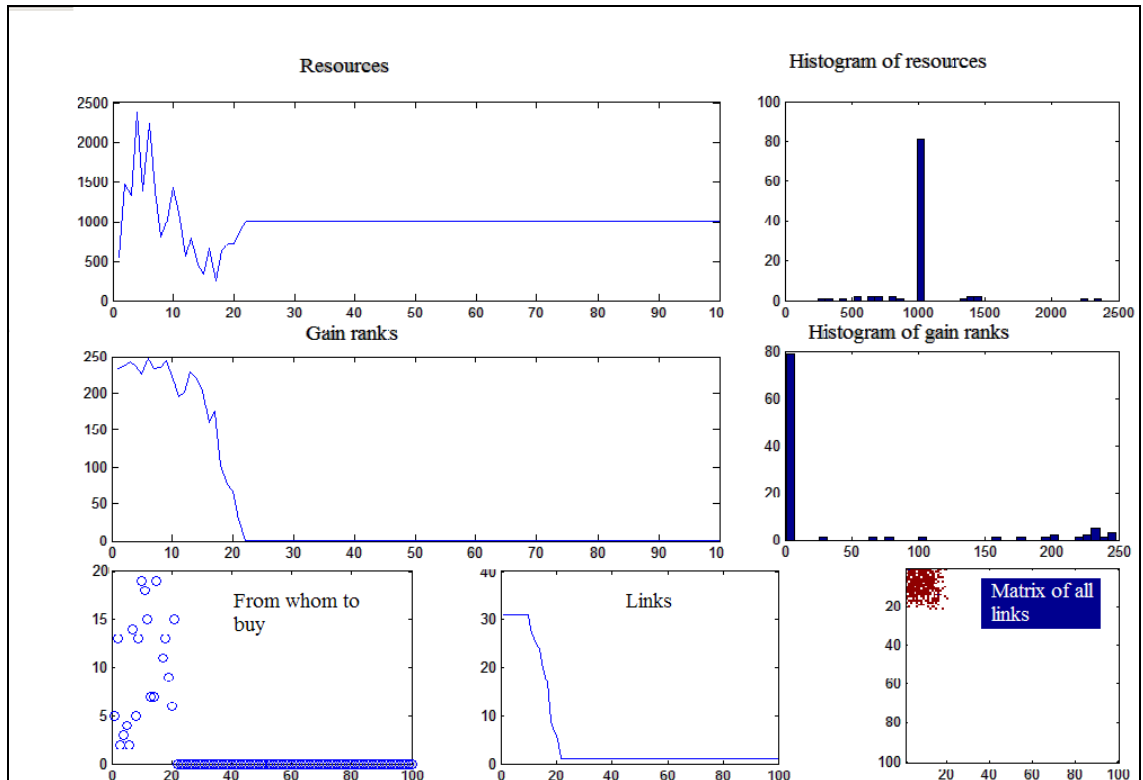

Figure 3a: An initial step of simulation where new agents join the exchange process.

Figure 3b depicts the next step of the link exchange process, when later joining agents start to get a return on their investment, i.e. their resources grow above the initial capital of 1000. More than 10 agents have more than their initial resources, including 6 agents who joined later; at the same time the resources of almost 20 agents were invested without any return. The rate of investment of agents which just joined is comparable with that of those who joined earlier. Regarding the resources histogram, the bar for agents with zero resources has started to grow.

The reader can see the spikes, which are down to zero on both the resources and gain rank curves. These agents ran out of resources and cannot afford to buy more links. Therefore they are unable to keep up with the other agents. Other than for these agents, the gain rank grows monotonically with the time spent in the exchange process. Two areas start to appear on the histogram of gain rank: for initial agents and for those who joined in the middle of the process in between the first and last agents.

The link matrix displays the preference of most agents to buy from the initial players; moreover, most of the agents tend not to buy from agents 30-40. There is an exception, clearly indicated on the buying choice graph: the new joiners buy from themselves since they cannot afford more expensive links, having their resources approaching zero.

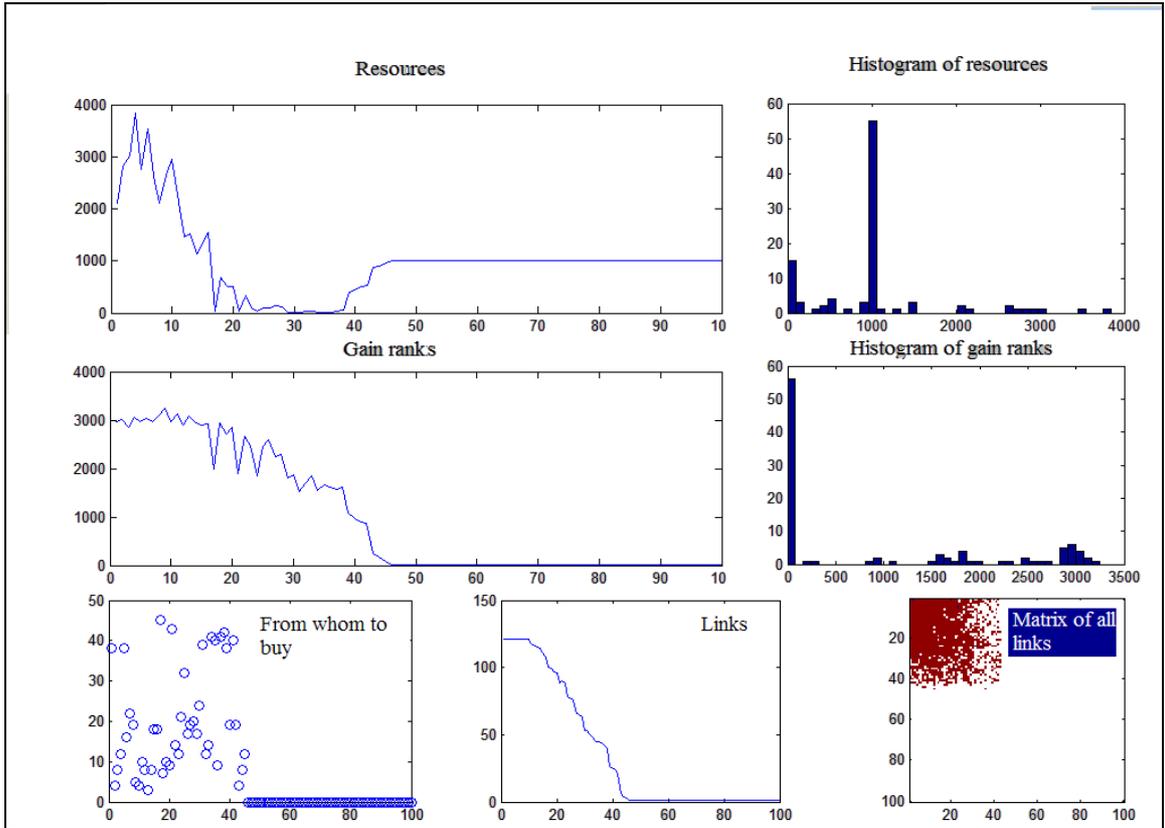

Figure 3b: Further step of simulation with joining new players

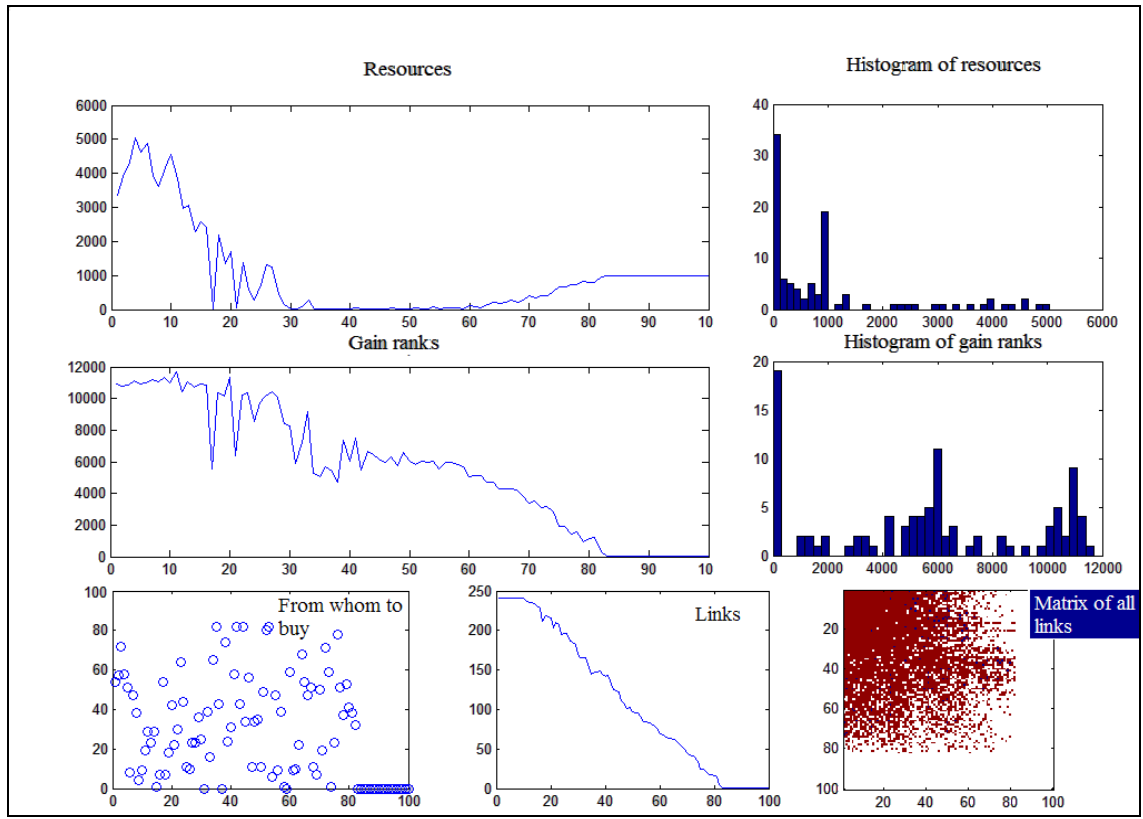

Figure 3c: Almost all agents have joined the game.

Figure 3c depicts the situation when almost all the agents have joined the game. A small portion of agents, who joined the process after start, have had positive returns (from agent 11, and approximately, to agent 25. Most of the resources are transferred from the agents, which just joined the game to ones who have joined much earlier. For a non-early agent who joins the game, it takes about 20 time steps (i.e. 20 new agents to join the game) for the resources of this agent to go down to zero. At the same time, it takes about 55 agents to join, for a given agent (who was not an initial agent) to bring its resources above the initial level (of 1000 units), if this agent did not have bad luck and its resources went down to zero anyway. The histogram of resources shows a high number of agents between the bars for zero and initial resources. Thereafter, it is almost uniformly distributed between the successful agents having up to 5 times the initial resources.

The gain rank of initial agents is high and almost the at the same level for all of them. The gain ranks of the next 30 agents which joined after that, have quite high volatility and decrease on average for agents who joined the game at a later step. The volatility for this group of agents is explained by fact that when some of them have less sales and therefore can only buy less affordable links, this further decreases their future sales capacity.

Towards the end of game the current step shows an almost uniform choice of sellers. The link matrix shows that only the agents with resources approaching zero (agents 30-50) buy from the agents which just joined (65-83); see right middle area in the link matrix. In contrast to the earlier steps of the process, the last of the joining agents prefer not to buy from themselves as they were inclined to do at the previous simulation steps. Furthermore, almost everyone among the first 40 agents has bought at least a single link within the group.

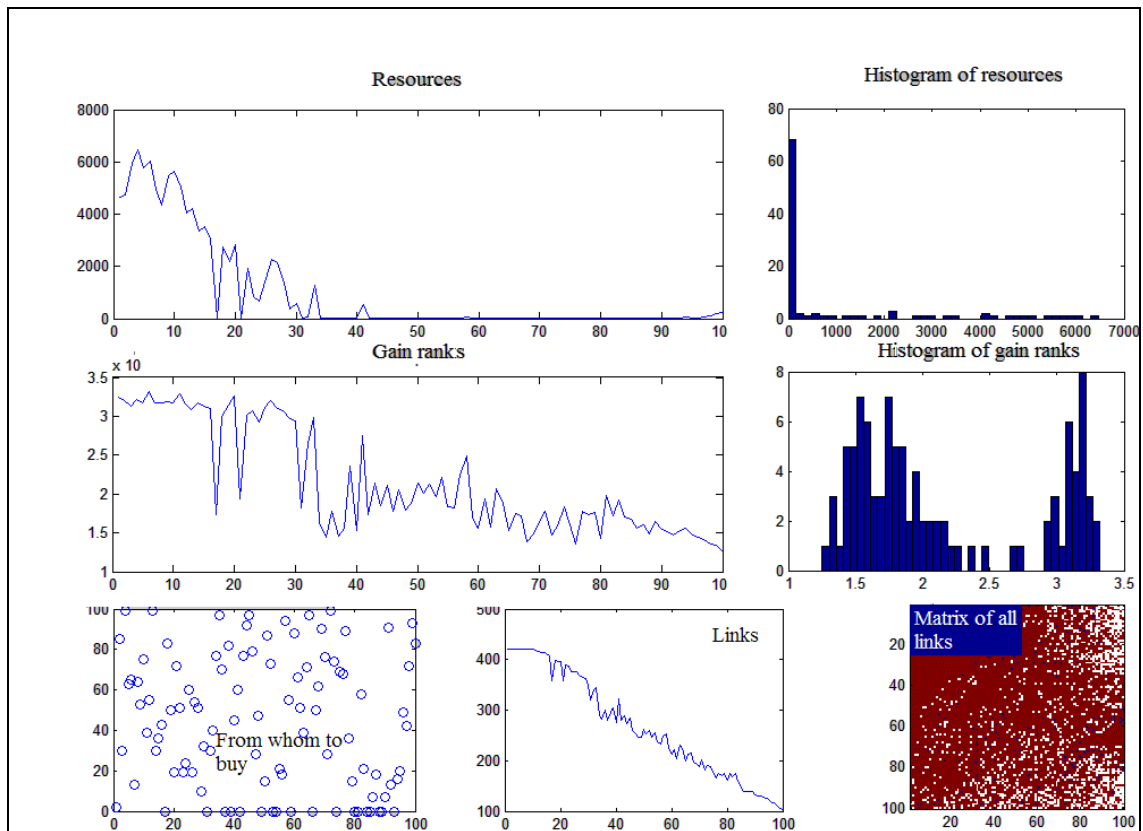

Figure 3d: Towards the end of the simulation run.

Figure 3d depicts the final simulation step, when all the agents have joined the process, almost all the links are bought and the distribution of resources has come to an equilibrium. Almost no additional agents, other than those who have joined the process at the early stages, have accumulated more resources than their initial allocation. All of the resources from the newly joining agents have been transferred to the initial set of agents. In terms of resources we have a 'winner takes all' result, and the distribution of the resultant resources is almost uniform, except that the majority of all agents have lost all of their resources.

There is a plateau for the gain rank for winners, followed by the two areas of lower value with high volatility for the later joining agents. The histogram of the gain rank is bi-modal: the right pyramid is formed by the rich agents, and the left is formed by those who lost all or most of their resources. Note that this represents two separate respective clusters of 'rich' and 'poor' agents with insignificant overlap.

At the equilibrium state the choice of buyers is uniform except that the majority of agents do not have resources to buy (zero value in the left bottom graph).

# Varying the rate of joining

In this section we consider the family of processes above obtained by varying rate of joining. The rate of joining is implemented via a probability of joining, 0<p<1, which is the probability that a new agent joins the process at a current step. In the above analysis we have considered the case where p=0.3. We will see that the parameters of the exchange process significantly depend on this rate of joining. We will subsequently choose the rate that provides the best fit with empirical observations from the real-world. We will refer to p<0.5 as a slow joining rate and p≥0.5 as a fast joining rate (we call p=0.5 the intermediate rate).

Under a slow joining process with p=0.1 (see Figure 4a), there is sufficient time for all the resources of newly joining agents to be transferred to the initial agents. There are zero resources for those that joined after the agent number 12, and everyone except the initial agents lost all their resources. There are distinct plateaus for the gain rank of the initial set of agents, for those who join shortly after that, and for the rest of the agents. We observe that there is a clear plateau for the gain rank for those which were last to join. The slow rate of joining gives us a clear 'winner takes all' scenario. Note that the resources of initial agents are distributed non-uniformly having 5 to 11 times of their initial resources.

The link matrix shows that those who have just joined tend not to buy from the initial set of agents but rather from themselves and from those who joined later. Initial agents do not buy from those who have joined some time ago and lost their resources (note two triangular areas in the link matrix on the right top and a larger area on the left bottom).

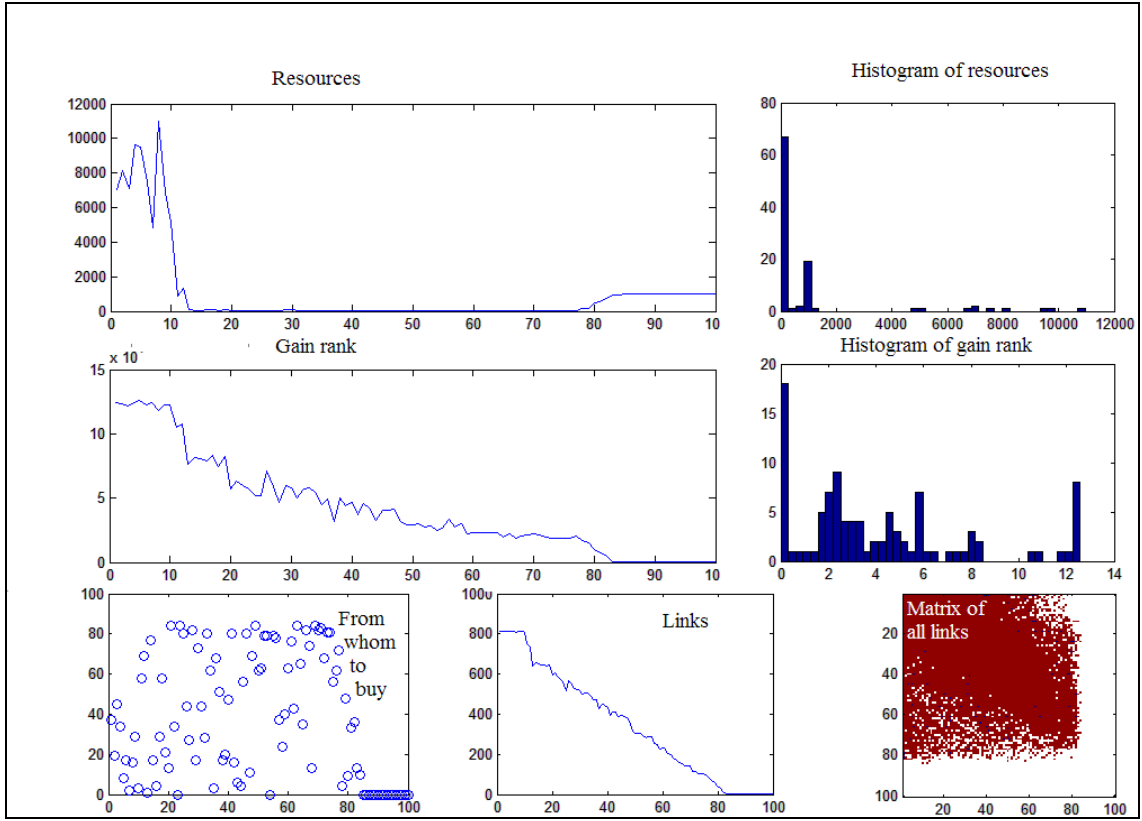

Figure 4a: Slow joining (p=0.1)

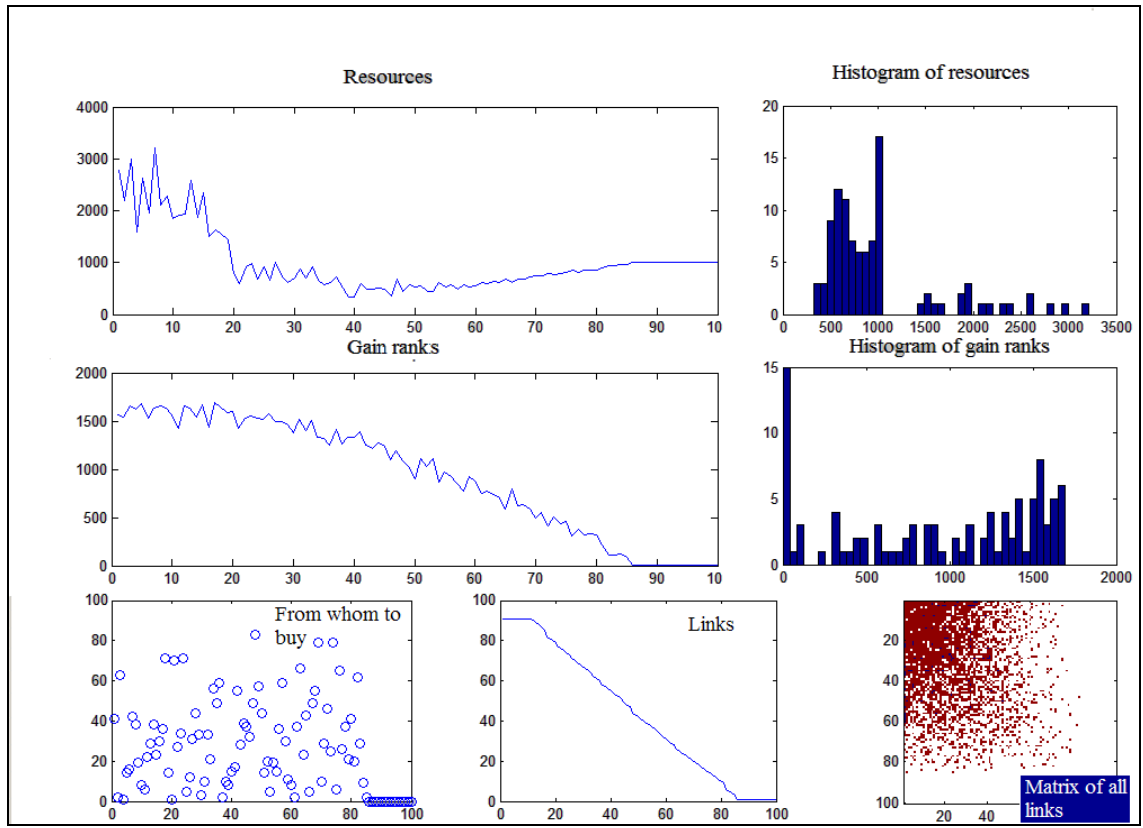

Figure 4b: Fast joining rate (p=0.9)

Conversely, under a fast joining process, with p=0.9 (see Figure 4b), the resource curve does not go all the way to zero and later joining agents gradually start to recover their investments. It turns out that also when p is high, the volatility of resources of initial agents is as high as under the slow joining process, but both the resources and gain rank curves are smoother in the fast joining process. We observe that the gain rank curve is monotonically decreasing. The longer agents are in the game, the slower they accumulate their links. The link curve itself is relatively linear and smooth. The newly joined agents tend to sell and buy links at a low volume that is seen in the link matrix. Therefore, the exchange process under a fast joining regime is quite different from that of the slow joining regime.

In contrast to the slow joining process, the fast joining process gives new agents a reasonable expectation to successfully gain a return on their investment. Therefore, the fast joining process can be considered as rational from the agents' perspective. However, on the macro-level, we choose the rate of the exchange process to provide the distribution curves that possess similar properties to those of the empirical curves, as shown in Figure1. Therefore, we heuristically choose the joining rate of p=0.5 (Fig. 4c) as a plausible model of reality.

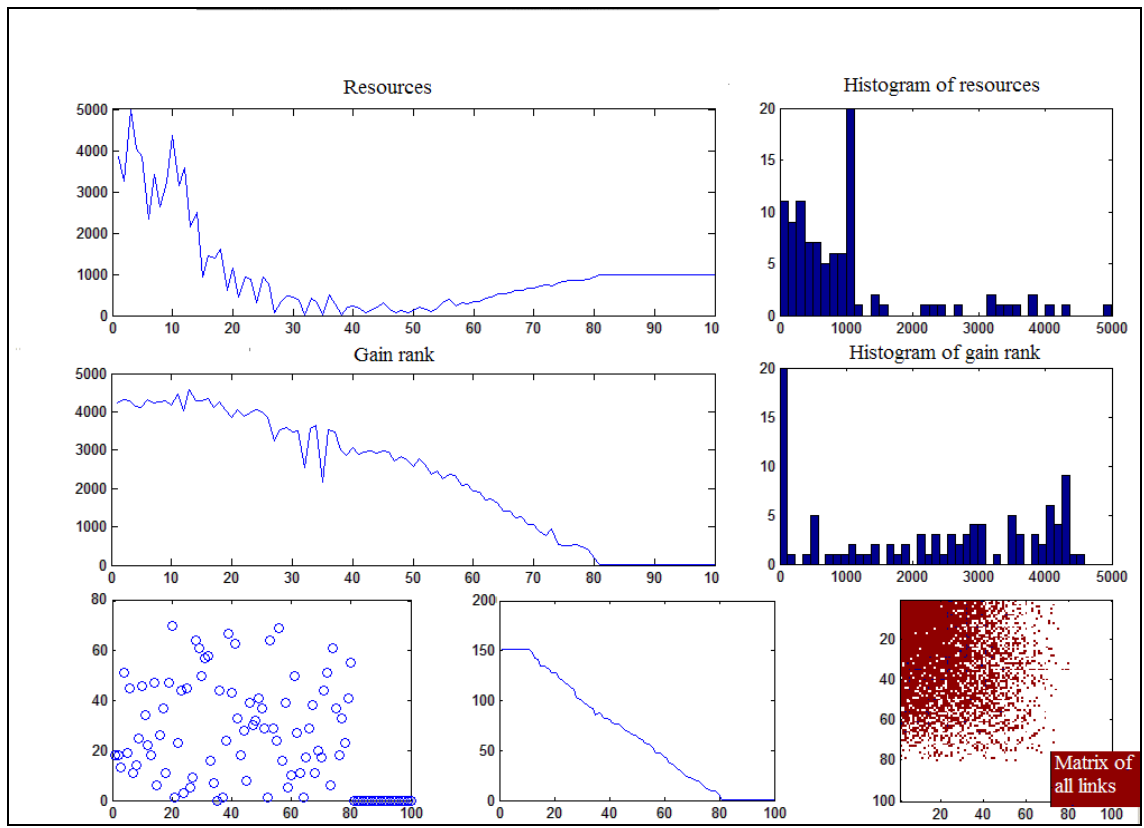

Figure 4c: Intermediate joining rate (the most plausible, p=0.5)

We have considered two cases. In the first case the agents join the game at a relatively slow rate and the initial agents have sufficient time to "pump out" all the resources from later joining agent (see Figure 5a below). As a result there is no benefit for the new agents to join the game at all. Conversely, if the rate of agents joining the game is fast enough so that the flow of resources from new agents to the initial agents is partially compensated by the flow of resources from the agents which have joined later to these new agents, then the business process becomes profitable for a certain portion of new agents that have not joined the game too late.

In the situation when the probability of joining is 0.5, then about 10% of agents are in the group of the winning initial agents, and about 15% of newer agents are able to gain more resources than they initially started with. The remaining 75% of agents lose almost all of their resources, but most of them still have a small amount of resources left.

As to the difference in the rate of growth for resources and gain rank, we observe the following under the slow joining regime (Figure 5a)

- Resources for initial agents show logarithmic growth with high initial volatility.

- The first joining agents start to grow their resources after an initial volatile period, when their resources were approaching zero.
- The other agents' resources are monotonically approaching zero, the earlier they joined the game the faster their resources approach zero.

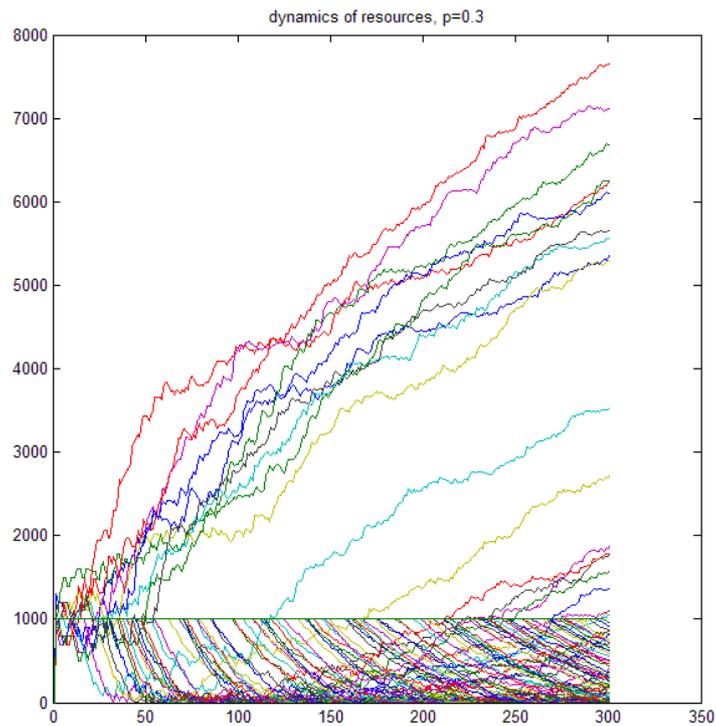

Figure 5a: Resources of agents versus time, for the slow joining process; the family of curves for each agent is depicted.

Under the fast joining regime (Figure 5b) we observe that:
- Resources grow logarithmically for the initially joined agents.
- Resources drop, approaching zero for the late joining agents

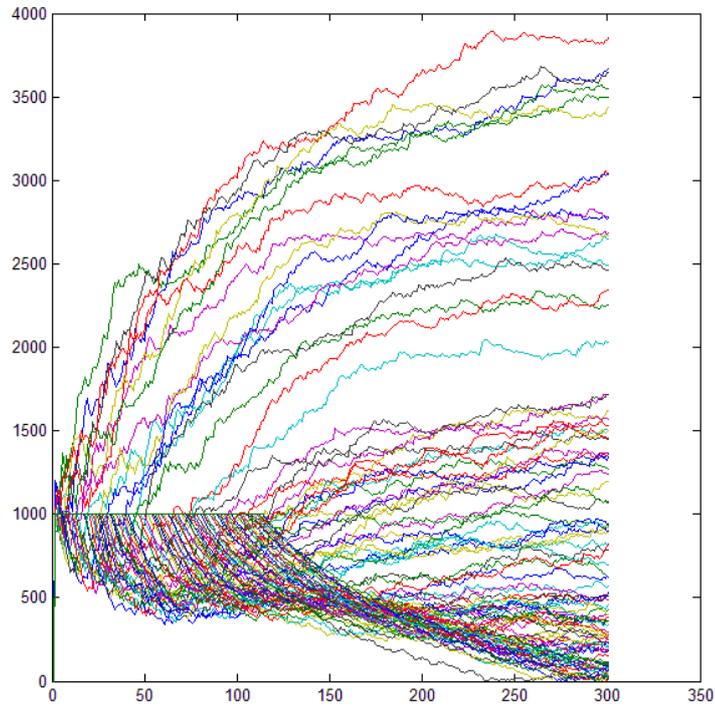

Figure 5b: Resources of agents versus time, for the fast joining process; the family of curves for each agent is depicted.

| p | Number of agents, apart from initial ones, who increased their resources as a result of the exchange process |
|---|---|
| 0 | Initial agents only |
| 0.1 | 6 |
| 0.2 | 7 |
| 0.3 | 8 |
| 0.4 | 10 |
| **0.5** | **12** |
| 0.6 | 13 |
| 0.7 | 14 |
| 0.8 | 16 |
| 0.9 | 17 |
| 1 | Distributed between all agents |

Table 1: The number of agents who successfully had a return on their investment (i.e. obtained higher than 1000 units), apart from the initial agents. The row presented in boldface, with p=0.5, is heuristically selected as the most plausible (see Figure 4b).

Table 1 shows the number of agents who joined the process and benefited from it. The number of such agents is considered to be a rationality measure. If no newly joining agents apart from the initial set of agents gain any additional resources, then the process is deemed to be irrational. Table 1 provides additional justification for our choice of p=0.5 as a reasonable rate for modeling reality. On the one hand, it gives rise to rational behaviour of agents, since 12 agents benefited in this case in addition to the 10 initial agents, and, on the other hand its resources histogram is realistic when heuristically compared to the empirical distributions of resources.

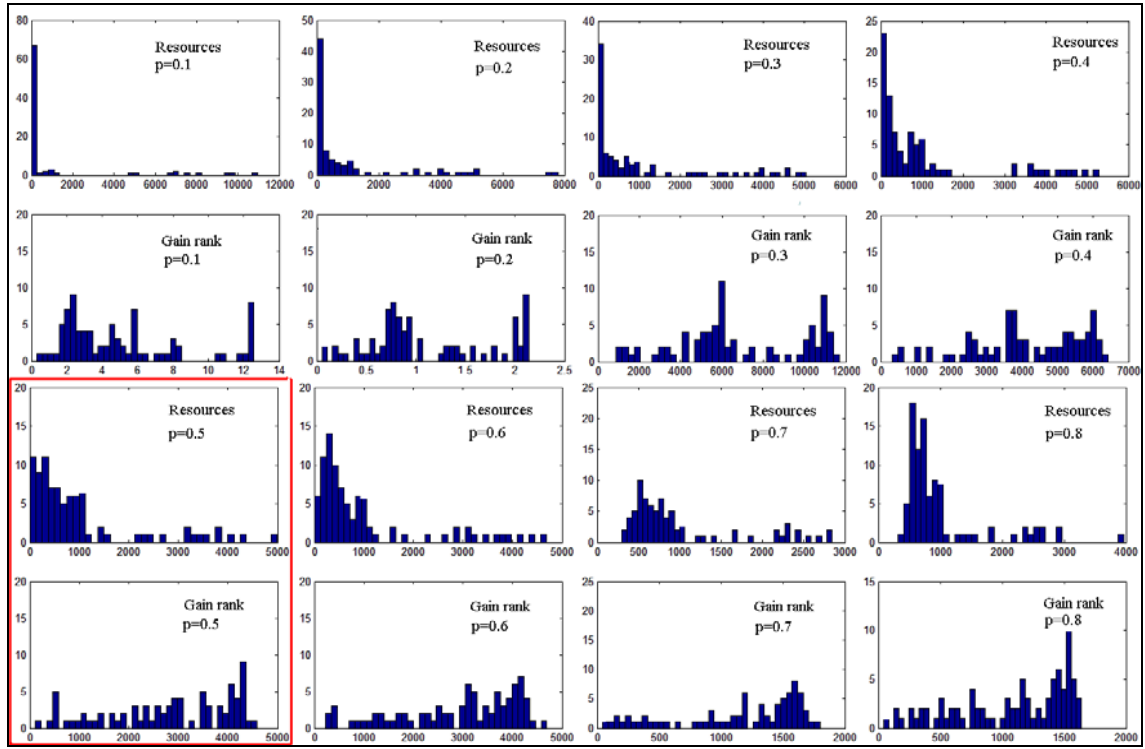

Figure 6: Resultant distributions of resources and gain ranks for various rates of joining. The most plausible simulation for p=0.5 is highlighted. Note that in contrast to Figures 2-4 we do not take into account the players which have not joined the game.

The resultant distributions for the values of joining probability p=0.1 to 0.8 are depicted at Figure 6. We observe that the power law phenomenon for resources is quite consistent with respect to the rate of joining and simulation step (the first row for slow joining (p=0.1 to 0.4) and the second row for fast joining (p=0.5 to 0.8)). At the same time, we observe that starting at p=0.6, the spike in the resources histogram (representing the highest number of agents) is nonzero. Such a situation is less plausible in terms of a power law distribution in comparison with the histograms for p=0.1 to 0.5.

Dramatic changes occur within the gain rank histogram, its distribution patterns are transformed from bimodal to unimodal as the rate of joining increases. Indeed, for the fast joining process, the histogram for the inverse value of gain rank is heuristically seen to follow a power law distribution. For the slow joining process, the left area on the histogram shows *poor* agents who lost their resources during earlier simulation steps and therefore do not achieve higher gain rank. On the contrary, the right area presents the *rich* agents who keep increasing their gain rank with time. Yet the difference between the gain rank of the rich and poor is proportionately smaller than their respective difference in resources. So, a small difference in gain rank may lead to a significant difference in resources.

In the real world, similar distributions are obtained for the investments means, where every agent invests approximately the same for the production means. As businesses develop, the value of these investments vary significantly (what we observe with the resultant accumulated resources).

## Simulation results

In this study we have simulated the process of web link exchange. In our artificial world web links are considered to be the means of investment. Agents buy links from other agents to themselves to increase their gain rank, and the return on investment is implemented via selling links to other buying agents, priced in accordance to the gain rank.

Our model demonstrates a wide range of possible outcomes in the link exchange process according to the rate at which agents join the game. The simulation trends show the dynamics of the beginning of the process when just a small number of links are sold (i.e. commercialised), and the further simulation steps when a large number of links are sold. The former corresponds to the current web economy, and the latter corresponds to the one future possibility for the economy of commercialised web links. The model allows tracking of the various contributions of commercialised links to the overall web topology and judging the resultant impact on the economy.

Our model shows the possibility of power law distributions of the resources and of the inverse of the investment means (gain rank). The agents in our simulation process demonstrate simple yet rational behaviour. We have shown that for certain parameter settings for the link exchange process and the rate of joining, power law distributions for resources are reproduced by the simulation, and that the gain rank distribution is consistent with our intuition when the sale of links is prevalent. Moreover, joining agents have rational expectations of increasing their resources in the link exchange process if a sufficient number of new agents join within certain time limits.

We observe that when the rate of joining is too fast p>0.5, the resources histogram tends to peak at nonzero values and therefore move away from strict power law behaviour. Under the slower rate of joining all the resources are transferred to the initial agents, which reproduce the natural market behaviour following the 'winner takes all' scenario, that is irrational. When no new agents join the exchange process, our simulation does not provide a realistic market scenario.

# Concluding Remarks

It is worth mentioning that even if a particular agent avoids to participate in the economy of links, the value of links in the ranking algorithms of search engines and other means of measuring importance on the web, make it impossible to entirely ignore the possibility of a web link economy (Walker 2002).

In this paper we have suggested a framework for a standardised exchange rate between links and conventional currencies. Affiliate programs and banner ads could be seen as establishing an exchange market, but these are based on more than the presence of a link. Most banner ads pay only for click-throughs and affiliate programs pay only when an agent who follows a link then makes a purchase. Whatever the mechanism for charging a customer for receiving a link, it is evident that links have already become an important investment instrument on the web.

Although at the present time open exchange of links for real-world money is relatively infrequent and highly discouraged by the search engines, there is a growing black market for links. Webmasters are already buying links from search engine optimisation companies that set up link farms and web rings to increase the PageRank of the recipient of a link. According to (Walker 2002), there is also a common law perception of selling one's integrity for links. If Google discovers such attempts to manipulate its PageRank, the site may be penalised by the assignment of a low or zero PageRank to the offending site. Also, there exist various forms of trade in this economy of links such as barter exchange. Reciprocal linking and link exchange are common practice, and are loosely arranged as favours or more systematically through web rings and blogrolling. Such actions may also affect Google's objective measurement of links (Hiler 2002).

The possibility to use commercialised links for estimating the relevance of documents is partially associated with the very nature of keyword search. Under standard keyword search, customers are expected to receive a long series of answers because of a lack of understanding accuracy. Under the *natural language question answering* model (Galitsky 2003), which is capable of delivering a more exact and accurate answer, PageRank may become less important and the relevance of a document would not rely on the structure of links, which is subject to commercialisation. Also, when search can be personalised to the individual user, the importance of link-based metrics such as PageRank will decrease, under the assumption that there is a manageable set of content-based relevant results for the user to inspect.

# About the Authors

Boris Galitsky is a lecturer at the <u>School of Computer Science and Information Systems</u> at Birkbeck College, University of London.

Web: http://www.dcs.bbk.ac.uk/~galitsky
E-mail: galitsky@dcs.bbk.ac.uk

Mark Levene is Professor of Computer Science in the School of Computer Science and Information Systems at Birkbeck College, University of London.
Web: http://www.dcs.bbk.ac.uk/~mark
E-mail: m.levene@dcs.bbk.ac.uk